\documentclass[prl,aps,showpacs,twocolumn,preprintnumbers,nofootinbib,amsmath,amssymb,superscriptaddress,longbibliography]{revtex4-2}
\usepackage{graphicx}
\usepackage{comment}
\usepackage[caption=false]{subfig}
\usepackage[normalem]{ulem} 
\usepackage{xcolor}         
\usepackage{soul}
\usepackage[T1]{fontenc}
\usepackage[english]{babel}
\usepackage{amsmath,amssymb,amsfonts}
\usepackage[colorlinks=True,linkcolor=red,citecolor=blue,urlcolor=blue]{hyperref}
\usepackage{bookmark}
\usepackage{cancel}
\usepackage{float}
\usepackage{xcolor}
\usepackage{chngcntr}
\usepackage{braket}
\usepackage{nicefrac}
\usepackage{chemmacros}

\newcolumntype{C}{>{$}c<{$}}

\begin{document}
\title{Finite-bias Coulomb blockade thermometry}

\author{Omid Sharifi Sedeh}
\email{omid.sharifisedeh@unibas.ch}
\affiliation{Department of Physics, University of Basel, Klingelbergstrasse 82, CH-4056, Basel, Switzerland} 

\author{Christian P. Scheller}
\affiliation{Department of Physics, University of Basel, Klingelbergstrasse 82, CH-4056, Basel, Switzerland} 

\author{Henok Weldeyesus}
\affiliation{Department of Physics, University of Basel, Klingelbergstrasse 82, CH-4056, Basel, Switzerland} 

\author{Taras Patlatiuk}
\affiliation{Department of Physics, University of Basel, Klingelbergstrasse 82, CH-4056, Basel, Switzerland}

\author{Matthias Gramich}
\affiliation{Pico group, QTF Centre of Excellence, Department of Applied Physics, Aalto University, P.O. Box 15100, FI-00076 Aalto, Finland} 

\author{Jukka P. Pekola}
\affiliation{Pico group, QTF Centre of Excellence, Department of Applied Physics, Aalto University, P.O. Box 15100, FI-00076 Aalto, Finland} 

\author{Dominik M. Zumbühl}
\email{dominik.zumbuhl@unibas.ch}
\affiliation{Department of Physics, University of Basel, Klingelbergstrasse 82, CH-4056, Basel, Switzerland} 

\date{\today}
\begin{abstract}
Coulomb blockade thermometers (CBTs) are versatile and, in principle, primary thermometers operating down to the micro-Kelvin range but bias heating spoils the thermometry and the primary mode. Here, we introduce a method to extract the CBT electron temperature in the presence of heat created by an arbitrary bias voltage, and without assumptions on the heat flow mechanisms. The charging energy is extracted with high precision and without any other knowledge, thus making true primary thermometry possible. The experiment also reveals a subtle dependence of the charging energy on phonon temperature below 100~mK likely due to the amorphous AlO$_x$ tunnel junctions.
\end{abstract}

\maketitle
\textit{Introduction}---Temperature is a key parameter in quantum devices, setting the coupling strength between subsystems and determining which degrees of freedom remain active as temperature is lowered while others ``freeze out''.  
Thermometry and controlled injection of heat into the subsystems can be used to elucidate the intricate thermodynamics and help create a model of the device heat flow. Coulomb Blockade Thermometers (CBTs) are relying on charging of an island to suppress conductance \cite{Averin1991,Grabert1992,WasshuberBook} and 
have been extensively studied both theoretically
\cite{Pekola1994,Hirvi1996,Yurttagul2021,Pekola2022a,Golubev1992,Golubev1997} and experimentally
\cite{Pekola2022,Casparis2012,Bradley2016,Bradley2017,Prunnila2010ExSituCBT}. They have proven to be a reliable tool for measuring electron temperatures down to a few milli-Kelvins and are valued for their robustness and simplicity. Additionally, CBTs have been successfully employed in adiabatic nuclear demagnetization experiments, achieving micro-Kelvin temperatures through combined on- and off-chip cooling techniques \cite{Palma2017OnOff,Samani2022,Yurttagul2019Indium,Sarsby2020,Jones2020}, opening new avenues for micro-Kelvin transport measurements.

\begin{figure}[t]
\centering
\includegraphics[width=\columnwidth]{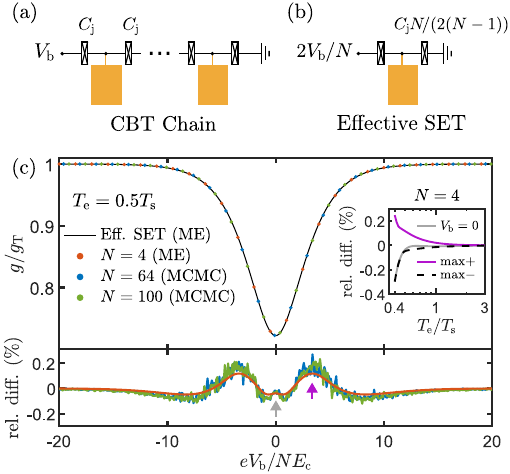}
\caption{
  Schematic of (a) a single-row CBT, and its (b) SET analogue. The orange rectangles represent metallic islands for thermalization to the substrate.
 (c) Normalized conductance \(g/g_{\rm T}\) is plotted against the scaled bias voltage \(eV_{\rm b} / N E_{\rm c}\) at \(T = 0.5T_{\rm s}\). The black curve represents the effective SET model, while the orange dots show the full numerical solution of ME for \(N = 4\). The blue and green dots correspond to MCMC simulations as labeled. For clarity, the upper panel shows only sparse simulation points, while the lower panel gives a full comparison. The relative difference shows excellent agreement with the SET analog at the 0.2\% level independent of N. Inset: Zero-bias and largest positive/negative deviations vs temperature. The bias voltage with the largest negative difference does not evolve continuously with $T_{\rm e}$.
}
\label{Figure1}
\end{figure}

Often, CBTs are realized in a chain of $N$ tunnel junctions with islands in-between, see Fig.~\ref{Figure1}(a), and $M$ identical chains in parallel. The chain with junctions in series provides resilience to voltage noise, and the parallel chains provide more current for measurement. In a voltage bias trace, CBTs exhibit a temperature-dependent conductance dip due to the Coulomb blockade effect with associated energy scale $E_c$, the charging energy. This suppression of conductance may provide the electron temperature $T_{\rm e}$ of the islands, which are often made from a metal. 

In the universal regime, CBTs are operating at \(T_{\rm e} \geq 0.4T_{\rm s}\), where $T_{\rm s}$ is a scaling temperature closely related to the charging energy $E_c$. In this regime, they are immune to offset charges and electrostatic disturbances, offering easy and robust thermometry \cite{Yurttagul2021,Samani2022}. At even higher temperatures \(T_{\rm e} \gg T_{\rm s}\), CBTs are serving as primary thermometers, allowing temperature extraction without relying on any other information. For this, the temperature is extracted from an approximation of the master equation (ME) fitted to the measured conductance versus bias curve. Equivalently, the linear proportionality of $T_{\rm e}$ to the full width at half minimum (FWHM) of the conductance dip can be used \cite{Pekola1994,Hirvi1996,Meschke2004}. Additionally, Markov Chain Monte Carlo (MCMC) simulations can extend the applicability of the primary-mode to lower temperatures, though at the cost of increased complexity and computational burden \cite{Samani2022,Yurttagul2021}.

These methods require finite bias, causing heating effects at low temperatures, thus generally spoiling the thermometry. Such heating effects may be taken into account in more sophisticated models \cite{Meschke2004,Casparis2012,Palma2017OnOff,Sarsby2020,Autti2023}, but the mechanisms of heat flow at low temperatures are often difficult to model and are themselves temperature dependent. For example, electron-phonon coupling with a \( T_{\rm e}^5 \) dependence may break down when the phonon wavelength exceeds the metallic island thickness or may become so weak that other heat flow channels become important. 
Thus, such modeling can be challenging and often not practical.
As a result, this has led to a preference for measuring the conductance at zero-bias where heating effects are minimal, but this mode requires calibration against another thermometer, thus losing the advantage of primary thermometry \cite{Palma2017OnOff,Samani2022,Yurttagul2019Indium,Sarsby2020}.

In this Letter, we present a new approach to CBTs operating as a primary thermometer in the universal regime in presence of heating from finite bias, and without assumptions about the thermalization mechanism. Based on a simple numerical solution of the master equation, we map the array onto a single island and extract the charging energy independently at each phonon temperature from the crossing point of two numerical solutions. This provides a precise calibration and thus constitutes a true primary thermometer, allowing extraction of the electron temperature $T_{\rm e}$ at any bias voltage. This bias voltage acts like a knob we can control to inject heat into the device and, combined with the thermometry, allows us to explore the thermodynamics of the device.
The measurements also exhibit a temperature dependence of the dielectric constant of the amorphous AlO$_{x}$ tunnel junctions with surprisingly strong change at the 10\% level below 100~mK. Our new method allows the study of heat flow and device parameters in an unprecedented way, opening new avenues for thermal analysis and versatile, reliable and precise thermometry at low temperatures.

\textit{Theory and Model}---In the absence of self-heating effects, the normalized conductance of a CBT in the high-temperature regime \( T_{\rm e} \gg T_{\rm s}(N) \) is well described by the first-order approximation of the master equation \cite{Pekola1994,Meschke2004,Hirvi1996}:
\begin{equation}\label{Eq1}
    \frac{g}{g_{\rm T}} = 1 - \frac{T_{\rm s}(N)}{T_{\rm e}} \mathcal{F} \left( \frac{eV_{\rm b}}{Nk_BT_{\rm e}} \right),
\end{equation}
where \( \mathcal{F}(x) = \left( x \sinh(x) - 4 \sinh^2(x/2) \right)/(8 \sinh^4(x/2)) \), describing the suppression of conductance due to Coulomb blockade. The high-bias conductance is \( g_{\rm T} = M/(N R_{\rm j}) \), \( R_{\rm j} \) the junction resistance, \( V_{\rm b} \) the bias voltage, \( k_B \) the Boltzmann constant, and \( e \) is the elementary charge.
The scaling temperature sets the depth of the dip and is given by \( T_{\rm s}(N) = \frac{2(N-1)}{N} \frac{e^2}{C_\Sigma k_{\rm B}} \), with \( C_\Sigma = 2C_{\rm j} + C_{\rm g} \) representing the total capacitance per island. Here, \( C_{\rm j} \) is the capacitance of each junction and \( C_{\rm g} \) is the island's capacitance to ground. In this study, we focus only on junction CBTs, where \( C_{\rm j} \gg C_{\rm g} \). However, the following method is also applicable to gate CBTs (\( C_{\rm g} \gg C_{\rm j} \)), as they behave almost identical to junction CBTs in the universal regime
\cite{Samani2022}. We emphasize that here the junctions in the chain are identical, and for a fixed $V_{\rm b}$ all islands are at temperature \( T_{\rm e} \). Further,
the charging energy \( E_{\rm c}=k_B T_{\rm s}/2 \) is assumed to be independent of \( V_{\rm b} \), and \( T_{\rm e} \). 

A CBT with arbitrary \( N \) behaves identically to a single-island CBT with \( N = 2 \) by substituting the effective capacitance \( C'_{\rm j} = \frac{C_{\rm j} N}{2(N-1)} \) and the partial voltage drop \( V' = 2V_{\rm b}/N \) in Eq.~\ref{Eq1}. This elegantly maps the chain to a single island CBT -- a single electron transistor (SET) -- with scaled parameters, as shown in Fig.~\ref{Figure1}(a), and (b). At lower temperatures, however, Eq.~1 breaks down, as it is only a first-order approximation, and the validity of the SET analogy needs to be verified. For this, we numerically solve the ME for a few-junction CBT using the damped simple iteration method \cite{Mokhlesi1997}. This technique relies on self-consistent computation of the probability distribution over possible charge states, but becomes computationally impractical as the number of configurations grows rapidly with $N$. Details are in the Supplementary Materials. 

In contrast, the MCMC method can be used to investigate much longer chains \cite{Hirvi1996,Yurttagul2021,Bakhvalov1989}. Rather than computing the entire probability distribution, MCMC simulates single-electron tunneling events stochastically, concentrating on the most probable configurations to obtain the conductance. 
We compare the two methods to the effective SET model, where we solve the ME for \( N = 2 \) with the scaled parameters. Figure~\ref{Figure1}(c) shows the excellent agreement within $\pm 0.2\%$ maximum error within the universal regime, validating the SET model as a computationally fast and accurate alternative to solving the full chain. Therefore, we use the effective SET model hereafter. 

\begin{figure}[t]
\centering
\includegraphics[width=0.93\columnwidth]{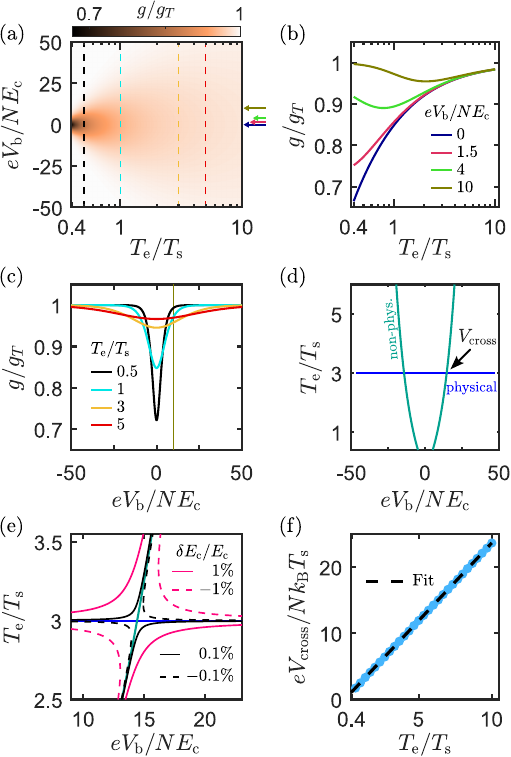}
\caption{
(a) $g/g_{\rm T}$ as a function of $eV_{\rm b}/N E_{\rm c}$ and $T_{\rm e}/T_{\rm s}$ with corresponding (b) horizontal, and (c) vertical cuts. 
(d) Conversion of the bias trace at $T_{\rm e} = 3T_{\rm s}$ (orange trace in (c)) into temperatures using the $g(V_{\rm b})$ dependence on $T_{\rm e}$. 
The conversion yields an expected physical solution of $T = 3T_{\rm s}$ (blue) and a non-physical (cyan) solution crossing at $\pm V_{\text{cross}}$. 
(e) Zoomed-in view of $V_{\text{cross}}$, showing anti-crossing behavior if the conversion uses a charging energy $E_c$ different from the actual value. 
The black and magenta curves correspond to the relative deviations of $\pm 0.1\%$, and $\pm 1\%$ in $E_{\rm c}$. 
(f) $V_{\text{cross}}$ as a function of $T_{\rm e}$. The fit yields $V_{\text{cross}} = (2.35 \pm 0.03) N k_B T_{\rm e}/e$.
}
\label{Figure2}
\end{figure}

The temperature dependence of conductance provides a means for electron thermometry under finite bias. This can be understood from the horizontal line cuts of Fig.~\ref{Figure2}(a), which reveal the temperature dependence of the normalized conductance at a fixed bias $V_b$ for a CBT with $N$ junctions. The line cut at zero bias in Fig.~\ref{Figure2}(b) demonstrates a one-to-one correspondence between conductance and electron temperature. Thus, with this conversion curve at hand, the experimentally measured conductance values can be directly converted to temperatures.

At finite bias, however, this one-to-one correspondence is breaking down. This is evident in Fig.~\ref{Figure2}(c), where the conductance dip becomes deeper and narrower as temperature decreases. At a fixed and sufficiently large bias — e.g., the vertical olive line — the conductance first decreases and then increases as the temperature is lowered.
This behavior is more clearly visible in Fig.~\ref{Figure2}(b), as shown by the olive curve, where a single conductance value corresponds to two different temperatures, leading to ambiguity when converting conductance into temperature. To resolve this, we first locate the minimum of the temperature trace and identify two possible solutions — one to the left and one to the right of the minimum.

Only one of the solutions is physical. By always choosing the physical solution, we can convert each conductance value at an arbitrary bias into an electron temperature. For example, the orange conductance trace in Fig.~\ref{Figure2}(c) can be converted into a temperature trace, see Fig.~\ref{Figure2}(d), by picking the physical (blue) solution over the non-physical one (cyan). This provides us with an electron temperature for each bias, in this simple case the same temperature at all bias values. In the experiment, this can be more complicated and contain more information e.g. due to heating effects, as discussed later. 

It's interesting to note that the two solutions intersect at a specific bias voltage -- the crossing voltage \( V_{\rm cross} \). At this voltage, the physical and non-physical solutions coincide. This crossing point turns out to be extremely sensitive to the value of the charging energy used for the conductance-temperature conversion. This is illustrated in Fig.~\ref{Figure2}(e), showing a zoomed-in view of the crossing point. Even a small deviation of only \( \pm 0.1\% \) in charging energy results in a significantly altered temperature curve, now exhibiting an anti-crossing at \( V_{\rm cross} \), see black curves. This sensitivity makes it a very accurate tool for extracting the charging energy \( E_{\rm c} \). The crossing voltage turns out to be defined by the electron temperature and the number of junctions in the array: \( \Delta V_{\text{cross}}\equiv2 V_{\text{cross}} = 4.7 N k_B T_{\rm e}/e \), as confirmed in Fig.~\ref{Figure2}(f) where the numerically extracted crossing point is shown over a range of electron temperatures. Notably, this is slightly lower than the voltage given by the known FWHM relation of the conductance dip, \( \Delta V = 5.439 N k_B T_{\rm e}/e \), which has been traditionally used as a primary thermometer.

This now makes possible a novel method to analyze experimental data for thermometry: We first extract the charging energy with the crossing voltage technique as detailed above. Then, each conductance point is converted to an electron temperature separately for each bias voltage, resulting in a full temperature trace as a function of bias voltage. This contains important information e.g. electron temperatures deviating from the lattice due to bias heating and allows to analyze the heat flow in the system.  

\begin{figure}[t]
\centering
\includegraphics[width=\columnwidth]{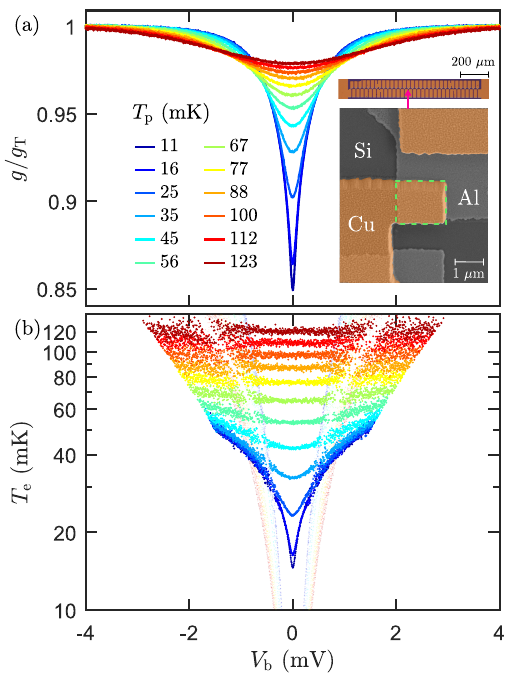}
\caption{
(a) Normalized conductance $g/g_{\rm T}$ vs. bias voltage $V_{\rm b}$ for Cu2, measured between $11 \, \text{mK} < T_{\rm p} < 123 \, \text{mK}$.
Inset: Optical image of a CBT row (top right) and a false-colored SEM image (bottom right) showing the position of an Al/AlO$_x$/Al tunnel junction at the overlap between the copper (Cu, orange) and aluminum (Al, light gray) layers, as highlighted by the dashed box.
(b) Color-coded $T_{\rm e}$ traces obtained by applying our method to the conductance traces in (a). The analysis is done for data points with more than $0.5\%$ conductance suppression. Blurred data points indicate non-physical solutions.
}
\label{figure3}
\end{figure}

\textit{Experimental implementation}---To demonstrate our method, measurements are performed on three CBT samples: two with copper thermalization islands (samples Cu1 and Cu2) with identical designs, each having 7 parallel chains of 64 junctions, and one with aluminum islands (labeled Al) having 10 chains of 100 junctions. All the junctions in these samples are Al/AlO$_x$/Al tunnel junctions.
In the copper samples, the junctions are located at the overlapping region of the copper (orange) and aluminum (light gray) layers (see Fig.~\ref{figure3}), and beneath the copper capping layer. This was done using the shadow evaporation technique, in which a 30 nm aluminum layer is first deposited at an oblique angle and oxidized in pure oxygen. Subsequently, a second aluminum layer, $\sim$30 nm thick, is deposited from a different angle to form the top electrode, which is then capped with a 150 nm copper layer. As a result, the islands in the copper samples have a total thickness of $210\ \text{nm}$. The lateral area alternates between $933 \ \mu\text{m}^2$ for even-numbered islands and $1'000 \ \mu\text{m}^2$ for odd-numbered islands.
In contrast, the aluminum CBT sample has its electrodes and islands fabricated entirely from aluminum, and the islands are $2 \, \mu\text{m}$ thick, with a lateral area of $1'190 \, \mu\text{m}^2$ (see supplementary materials).

The samples are glued into separate metallic boxes, each of which is thermalized and bolted to the sample stage of the dilution refrigerator. No filters were used on the sample wires, except for CBT1, which includes a pair of silver epoxy filters \cite{Scheller2014SilverEpoxy} located on the sample box. The sample stage has a RuO$_x$ thermometer, which is assumed to accurately reflect the phonon temperature $T_{\rm p}$ of the samples. A perpendicular magnetic field of 40 mT is applied to the samples to suppress the superconductivity in the aluminum layers.

The devices are measured in the same cool-down from $T_{\rm p} = 123 \, \text{mK}$ down to the base temperature of $\sim 10 \, \text{mK}$; see the measured data in Fig.~\ref{figure3}(a) and the Supplementary Material for other samples. After each $T_{\rm p}$ set-point, approximately one hour is allowed to ensure temperature stabilization and proper sample equilibration.  
Fitting Eq.~\ref{Eq1} to the three highest-temperature traces, where overheating effects are minimal, allows us to extract the high bias conductance $g_{\rm T}$, which was assumed as a common fit parameter. Using our new method, we extract the charging energy with the crossing point method and then convert the conductance trace into a temperature trace at each phonon temperature, as shown in Fig.~\ref{figure3}(b) and S2. At high refrigerator temperature, $T_{\rm e}$ remains constant over the bias range, while at lower refrigerator temperatures, self-heating effects become significant, causing $T_{\rm e}$ to rise with increasing bias, see the V-shaped blue curves.

\begin{figure}[t]
\centering
\includegraphics[width=0.90\columnwidth]{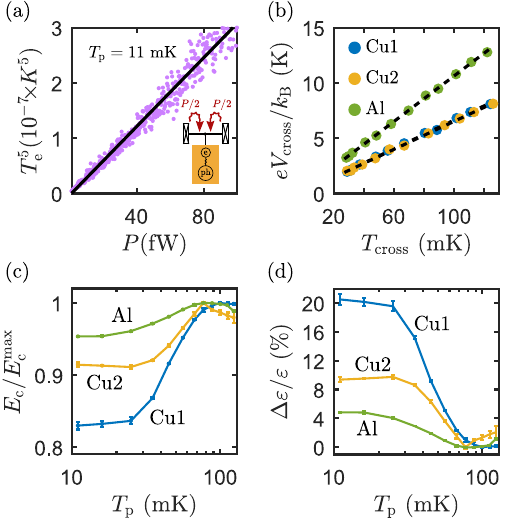}
\caption{
(a) $T_{\rm e}^5$ vs. power $P$, highlighting electron-phonon coupling as a bottleneck for electron thermalization in metallic islands at low temperatures. 
(b) $eV_{\rm cross}/k_B$ vs. $T_{\rm cross}$ shows a linear dependence, in agreement with our theoretical model. The fit yields \( N = 65 \pm 2 \), \( 64 \pm 2 \), and \( 103 \pm 2\) for Cu1, Cu2, and Al CBTs, respectively, matching the lithographic junction count of 64 and 100, respectively.
(c) Extracted normalized charging energy $E_{\rm c}/E_{\rm c}^{\rm max}$ and their corresponding (d) relative permittivity change $\Delta \epsilon / \epsilon$ as a function of $T_p$. Here,  $E_{\rm c}^{\rm max}/k_{\rm B} = 8.4$, $7.9$, and $21.9 \, \text{mK}$, for Cu1, Cu2, and Al samples, respectively.
}
\label{fig:Figure4}
\end{figure}

To investigate the cooling mechanism, we analyze the temperature data as a function of the dissipated power \( P = (V_{\rm b}/N)^2 / R_{\rm j} \) \cite{Kautz1992} at the lowest phonon temperature.  When plotting $T_{\rm e}^5$ the data falls on top of a line, as shown in Fig.~\ref{fig:Figure4}(a), thus indicating electron-phonon coupling as the cooling mechanism, consistent with previous studies \cite{Meschke2004, Meschke2011} for similar CBTs. We emphasize that our temperature extraction method confirms this behavior without relying on any assumptions about heat exchange mechanisms or specific thermal models. Moreover, it would also work for other cooling mechanisms, which might give exponents other than 5. Such investigations will now be possible for CBT devices using our primary thermometry method and the bias voltage as the heat-knob.
Finally, agreement with a clean power-law at a fixed phonon temperature also confirms the independence of the charging energy on the electron temperature, see below. 

The crossing points provide a method to verify how many junctions in the sample are functioning properly. To illustrate, we extract the crossing voltages, \( V_{\text{cross}} \), and their corresponding temperatures, \( T_{\text{cross}} \), from Fig.~\ref{figure3}(b) and plot them in Fig.~\ref{fig:Figure4}(b). We observe a linear relationship, agreeing with our derived relation, \( V_{\text{cross}} = 2.35 N k_B T_{\rm e} / e \). Using this relation and performing linear fits, we determine \( N \), which closely matches the fabricated number of junctions. Notably, this method can be used to detect shorted junctions (within the error bars), and remains effective even in the presence of self-heating effects. The FWHM of the conductance dip is also proportional to $N$, but becomes much narrower in overheated devices and then does not reflect $N$ anymore.

The analysis performed above in Fig.~\ref{figure3} reveals a rather surprising dependence of the charging energy on the phonon temperature for all three CBTs, see Fig.~\ref{fig:Figure4}(c). We note the relatively strong temperature dependence of $E_{\rm c}$ ranging from 5\% to $\sim15\%$ for the different CBTs, which is far larger than the $\sim1\%$ error bar on $E_c$. 
To explore the origin of the temperature dependence, the extracted charging energy values are converted into the relative change in the dielectric permittivity, \( \Delta \epsilon / \epsilon \), of the AlO$_x$ layer in the tunnel junctions using the relation \( E_{\rm c} = \frac{N-1}{N} \frac{e^2}{C_\Sigma} \). 
The resulting temperature dependence of the relative permittivity, shown in Fig.~\ref{fig:Figure4}(d), is reminiscent of the behavior of two-level systems (TLS) in amorphous materials at low temperatures \cite{Phillips1987Dielec,Frossati1977Dielec,Walker1984Dielec,Churkin2021Dielec}.
Previously, a similar TLS behavior was observed in AlO$_x$ oxide layers \cite{Fritz2018}, where $\Delta \epsilon / \epsilon$ exhibits a minimum around 100 to 150 mK, followed by a linear increase on a logarithmic scale, reflecting a $\ln(1/T)$ dependence. In our case, however, the change of the relative permittivity is two orders of magnitude larger. This could be due to large strain effects in our very thin oxide of only $\sim1$~nm, which are much thinner than the $\sim 20-30$~nm oxide layers of Ref.~\cite{Fritz2018}. Additionally, the shape of the temperature dependence we observe agrees with theory of strained amorphous materials \cite{Stockburger1995}.
Further investigation, such as varying the measurement frequency \cite{Frossati1977Dielec,Walker1984Dielec,Osheroff1996Dielec}, adjusting the oxide thickness, introducing deliberate mechanical strain \cite{Grabovskij2012}, or annealing techniques \cite{Pappas2024}, may provide additional insights into this observation in the future studies.

\textit{Conclusion}---Using an effective single-island model for the CBT chain, we obtain the charging energy with high precision from a bias sweep despite self-heating. This calibrates the CBT and allows extraction of the electron temperature at arbitrary bias voltages and without assumptions on the heat flow mechanisms, thus establishing a true primary thermometer. When varying the phonon temperature, a change of the charging energy is observed below $\sim100$~mK. This behavior is qualitatively similar to experiments on thicker oxide layers \cite{Fritz2018}, but is about two orders of magnitude larger here, possibly related to the very thin and highly strained amorphous tunnel junction oxide.

With these advances, applications in precise, fast and primary thermometry can be pursued. The charging energy can be extracted in situ, without requiring prior knowledge, providing a precise calibration with error well below 1\%. Once calibrated, the electron temperature can be extracted with just one measurement of CBT conductance, making possible fast, time-resolved temperature measurements. Further, one may use this method to study ultra-low temperature thermodynamics, such as energy exchange between different reservoirs or coupling between different degrees of freedom. In addition, accurate capacitance measurements could be done by depositing dielectrics on the CBT islands and capping them with a top gate. The resulting change in capacitance can be measured via the extraction of charging energy. This approach avoids parasitic capacitances appearing in standard CV measurements and supports studies at milli- and micro-Kelvin temperatures.


\textit{Acknowledgement}---
We thank Dr. Monica S. Schönenberger, Michael Steinacher, and Sascha Martin for their technical support.
This research was supported by the EU Horizon 2020 program through the European Microkelvin Platform (EMP) under Grant No. 824109 and the Energy Filtering Non-Equilibrium Devices (EFINED) project under Grant Agreement No. 766853. Additional funding was provided by the MSCA Cofund Action Quantum Science and Technologies at the European Campus (QUSTEC) under Grant No. 847471, the Swiss National Science Foundation (Grant No. 179024 and 215757), the Swiss Nanoscience Institute, and the Georg H. Endress Foundation.


%

\end{document}